\documentclass[apjl]{emulateapj}
\usepackage{epsfig}
\begin{document}

\title{On the formation of galactic thick disks}

\author{I.~Minchev\altaffilmark{1}, M.~Martig\altaffilmark{2}, D.~Streich\altaffilmark{1}, C.~Scannapieco\altaffilmark{1}, R.~S.~de~Jong\altaffilmark{1}, and M.~Steinmetz\altaffilmark{1}
}

\altaffiltext{1}{Leibniz-Institut f\"{u}r Astrophysik Potsdam (AIP), An der Sternwarte 16, D-14482, Potsdam, Germany}
\altaffiltext{2}{Max-Planck-Institut f\"{ur} Astronomie, K\"{o}nigstuhl 17, D-69117 Heidelberg, Germany}

\begin{abstract}
Recent spectroscopic observations in the Milky Way suggest that the chemically defined thick disk (stars with high [$\alpha$/Fe] ratios and thus old) has a significantly smaller scale-length than the thin disk. This is in apparent contradiction with observations of external edge-on galaxies, where the thin and thick components have comparable scale-lengths. Moreover, while observed disks do not flare (scale-height does not increase with radius), numerical simulations suggest that disk flaring is unavoidable, resulting from both environmental effects and secular evolution. Here we address these problems by studying two different suites of simulated galactic disks formed in the cosmological context. We show that the scale-heights of coeval populations always increase with radius. However, the total population can be decomposed morphologically into thin and thick disks, which do not flare. We relate this to the disk inside-out formation, where younger populations have increasingly larger scale-lengths and flare at progressively larger radii. In this new picture, thick disks are composed of the imbedded flares of mono-age stellar populations. Assuming that disks form inside out, we predict that morphologically defined thick disks must show a decrease in age (or [$\alpha$/Fe] ratios) with radius and that coeval populations should always flare. This also explains the observed inversion in the metallicity and [$\alpha$/Fe] gradients for stars away from the disk midplane in the Milky Way. The results of this work are directly linked to, and can be seen as evidence of, inside-out disk growth.
\end{abstract}


\section{Introduction}

The formation of galactic thick disks has been an important topic ever since their discovery in external galaxies \citep{burstein79, tsikoudi79} and in the Milky Way \citep{gilmore83}. A number of different mechanisms have been proposed for the formation of thick disks (see Introduction in \citealt{minchev12b} and references therein), but it is generally believed that they are the oldest disk components.

Stellar disk density decomposition into thinner and thicker components in external edge-on galaxies have found that the thin and thick disk components have comparable scale-lengths (e.g., \citealt{yoachim06, pohlen07, comeron12}). While this is consistent with results for the Milky Way when similar morphologically (or structural) definition for the thick disk is used (e.g., \citealt{robin96, ojha01, juric08}), it is in contradiction with the significantly more centrally concentrated older or [$\alpha$/Fe]-enhanced stellar populations (e.g., \citealt{bensby11, cheng12b, bovy12a}). This apparent discrepancy may be related to the different definition of thick disks - morphological decomposition or separation in chemistry.

Moreover, while observed disks do not flare \citep{vanderkruit82, degrijs98, comeron11}, numerical simulations suggest that flaring cannot be avoided due to a range of different dynamical effects. The largest source is probably satellite-disk interactions (e.g., \citealt{bournaud09, kazantzidis08, villalobos08}) shown to increase an initially constant scale-height by up to a factor of $\sim10$ in 3-4 disk scale-lengths.
Studying preassembled N-body disks, \cite{minchev12b} showed that purely secular evolution (in the absence of external perturbations) also causes flared disks, due to the redistribution of disk angular momentum. This was linked to the accumulation of outward/inward migrators in the outer/inner disk (with vertical actions larger/smaller than those of non-migrators), which caused vertical thickening/contraction and thus flaring.
In addition to merger perturbations and radial migration, misaligned gas infall \citep{scannapieco09, roskar10} and reorientation of the disk rotation axis \citep{aumer13a} can also produce disk flaring. 

In this Letter we show that the above outlined contradictions can be reconciled when realistic galactic disks growing in a cosmological environment are considered.
\begin{figure}
\epsscale{1.05}
\plotone{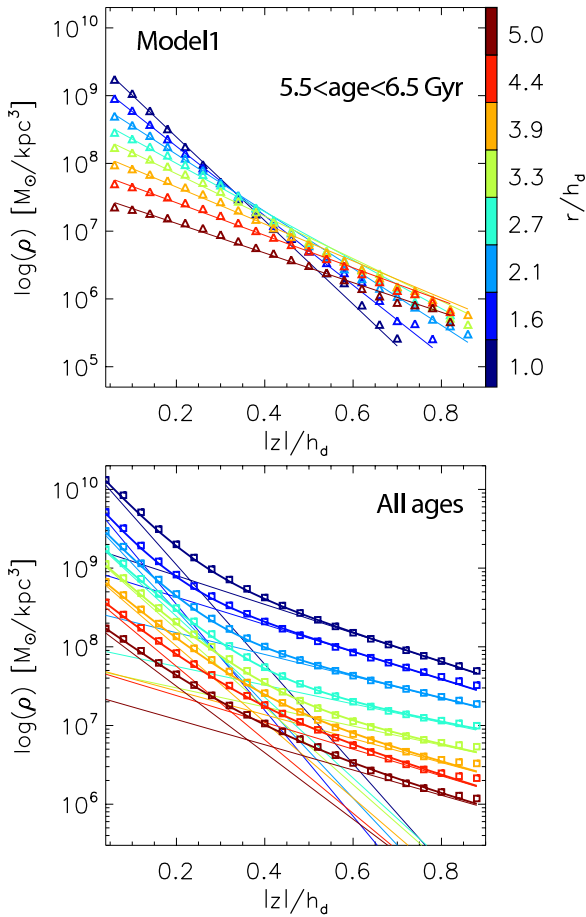}
\caption{
Scale-height fits for Model1. {\bf Top:} The vertical density profiles of mono-age populations (a typical one shown here) can be fitted well by single exponentials at all radii. Different colors correspond to different radial bins, as indicated by the color-bar. Distances are in units of the disk scale-length, $h_d$.
{\bf Bottom:} When the total stellar population is considered, for all radii a sum of two exponentials is required for a proper fit. Model2 gives similar results.
}
\label{fig:1}
\end{figure}
\newline
\section{Simulations}

We study two simulations following the formation of disks in the cosmological context, which use two distinct simulation techniques. 

The first model, Model1, is a gas dynamical simulation in a cosmological context using a sticky particle algorithm, as described in \cite{martig09, martig12}, with 150~pc spatial and 10$^{4-5}$~M$_{\odot}$ mass resolution. This is a barred late-type galaxy with stellar mass $4.3\times10^{10} M_{\odot}$ and a disk scale-length $h_d=5$~kpc, estimated from all stars in the range $4<r<12$~kpc, $|z|<2$~kpc. Further details about this simulation can be found in \cite{martig14a} (model g106) and \cite{mcm13}.

\begin{figure*}
\epsscale{1.}
\plotone{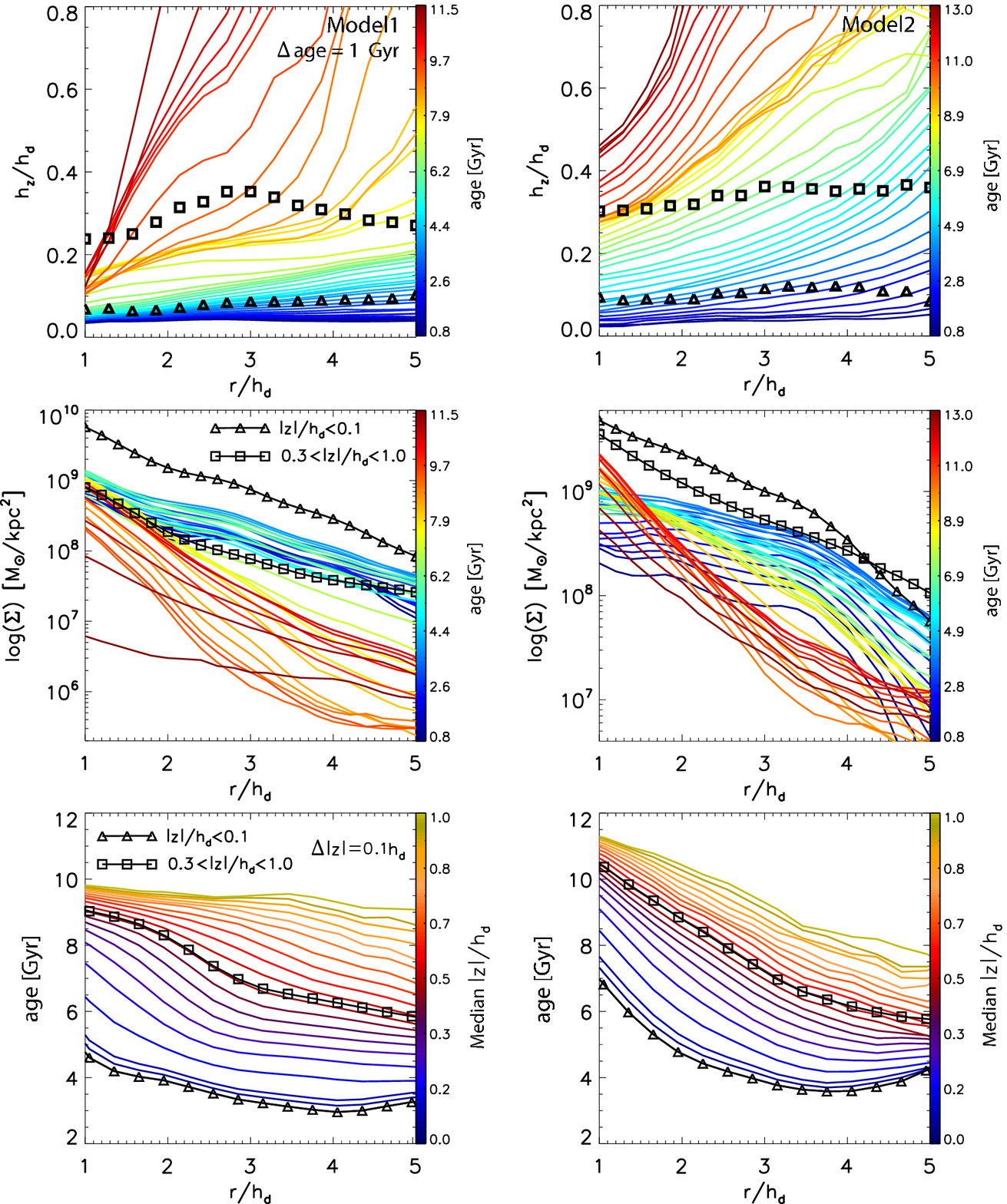}
\caption{
{\bf Top:} Variation of disk scale-height, $h_z$, with galactic radius for Model1 (left) and Model2 (right). Color lines show mono-age populations, as indicated. Overlapping bins of width $\Delta$age $=1$~Gyr are used. Overlaid also are the thin (triangles) and thick (squares) disks obtained by fitting a sum of two exponentials to stars of all ages (as in Fig.~\ref{fig:1}, bottom). No significant flaring is found for the thin and thick disks. 
{\bf Middle:} Disk surface density radial profiles of mono-age populations. Older disks are more centrally concentrated, which explains why flaring diminishes in the total population. Also shown are the surface density profiles of stars close to (triangles) and high above (squares) the disk midplane. The thickened disk component extends farther out than the thin one.
{\bf Bottom:} Variation of mean age with radius for samples at different distance from the disk midplane, as indicated by the color bars. Slices in $|z|$ have thickness $\Delta|z|=0.1h_d$. Overlaid are also the age radial profiles of stars close to (triangles) and high above (squares) the disk midplane. Age gradients are predicted for both the (morphologically defined) thin and thick disks. 
}
\label{fig:2}
\end{figure*}

The second model, Model2, is a full cosmological zoom-in hydro simulation, using initial conditions from one of the Aquarius Project haloes \citep{springel08, scannapieco09}. The technique used here is Tree-PM SPH with 300~pc spatial and $4.4\times10^5$~M$_{\odot}$ mass resolution. Model2 is a non-barred galaxy with stellar mass $5.5\times10^{10} M_{\odot}$ and a disk scale-length $h_d=4$~kpc, estimated in the same spatial range as for Model1. Further details about this simulation can be found in \cite{aumer13b}, their model Aq-D-5.

Both simulations form initial central components during an early epoch of violent merger activity. Gas-rich mergers supply the initial reservoir of gas at high redshift and merger activity decreases with redshift, similarly to what is expected for the Milky Way. This inside-out disk formation results in centrally concentrated older stellar populations for both simulations (see \citealt{martig14a} and \citealt{aumer14}). The general formation and evolutionary behavior of our disks is similar to many recent simulations in the cosmological context (e.g., \citealt{brook12, stinson13}).

We work in units of disk scale-lengths, $h_d$, which are estimated using the total stellar population at the final simulation time.

\section{Results}

\subsection{To flare or not to flare}

Similarly to \cite{martig14a}, we decomposed the stellar disks into mono-age populations, i.e., narrow bins of age, where we used $\Delta$age $=1$~Gyr. It was found that single exponentials provided good fits to the column density in the vertical direction for all age bins and both Model1 and Model2, in agreement with \cite{martig14a}. Assuming that chemistry can be a proxy for age, this is also consistent with studies of mono-abundance populations in the Milky Way \citep{bovy12a}. The top panel of Fig.~\ref{fig:1} shows single exponential fits for a typical mono-age population of Model1, in eight radial bins in the range $1<r<5$ disk scale-lengths, $h_d$. Radial bins of width $0.6h_d$ were used. It can be seen that the fits are very good for all radii. In contrast, the bottom panel of Fig.~\ref{fig:1} shows that a sum of two exponentials is required for a proper fit when considering the total stellar population.

It is clear already from the top panel of Fig.~\ref{fig:1} that significant flaring exists for the particular mono-age population shown (i.e., the vertical density profile flattens with increasing radius). However, only milder variations with radius are seen for the total population, when decomposed into thin and thick disks (Fig.~\ref{fig:1}, bottom panel). 

In the top row of Fig.~\ref{fig:2} we plot the scale-height variation with galactocentric radius, $r$, in the region 1-5 disk scale-lengths, $h_d$, for both Model1 (left panel) and Model2 (right panel). Forty overlapping mono-age populations of width 1~Gyr are shown, covering the age range in each disk, as indicated in the color bar on the right of each panel. Both the radius and scale-height, $h_z$, are in units of $h_d$. It can be seen that, indeed, for both models significant flaring is present, which increases for older coeval populations. For Model1 the increase in scale-height with radius for stars of age $\sim8$~Gyr is about a factor of four in four scale-lengths. The flaring decreases by about a factor of two for stars with age $<2$~Gyr. Even more significant flaring is seen for Model2.

In contrast to the flaring found for all mono-age populations, the thin and thick disk decomposition of the total stellar population including all ages, results in no apparent flaring. This is shown by the triangle and square symbols overlaid in the top row of Fig.~\ref{fig:2}.

We discuss the interpretation and implications of this interesting result in the following sections.

\subsection{Interpretation}

What is the reason for the flaring of mono-age disks? In numerical simulations flaring is expected to result from a number of mechanisms related to galactic evolution in a cosmological context (e.g., \citealt{bournaud09, kazantzidis08, villalobos08, aumer13a}). Even in the absence of environmental effects, flaring is unavoidable due to secular evolution alone (radial migration caused by spirals and/or a central bar, \citealt{minchev12b}). It should be stressed here that, while migration flares disks in the lack of external perturbations, during satellite-disk interactions it works {\it against} disk flaring \citep{mcm14}. Yet, this is not sufficient to completely suppress the flaring induced by orbiting satellites, as evident from the top row of Fig.~\ref{fig:2}. This suggests that external effects are much more important for the disk flaring we see in the simulations. Because the mass and intensity of orbiting satellites generally decreases with decreasing redshift, so does the flaring induced. It can be expected that at a certain time secular evolution takes over the effect of external perturbations.\footnote{\cite{minchev14} suggested that this time could also be inferred from the shape of the [$\alpha$/Fe]-velocity dispersion relation of narrow metallicity samples.}

Resolution and sub-grid models for gas physics may set an unrealistically high lower limit for the vertical velocity dispersion of young stars (e.g., \citealt{hause11}). The fact that the youngest populations in our simulations do not flare (dark-blue curves in top panels of Fig.~\ref{fig:2}) seems to indicate that numerical issues are not important for our result (see also \citealt{martig14b}, Sec.~5).

What is the reason for the lack of flaring in the total disk population? In an inside-out formation scenario, the outer disk edge (where flaring is induced) moves progressively from smaller to larger radii, because of the continuous formation of new stars in disk subpopulations of increasing scale-length. At the same time the frequency and masses of perturbing satellites decreases. Because of the inside-out disk growth, which results in more centrally concentrated older samples (see Fig.~\ref{fig:2}, middle row), the younger the stellar population, the further out it dominates in terms of stellar mass. Therefore, the thickened disk component results from the imbedded flares of different coeval populations, as seen in the top row of Fig.~\ref{fig:2}. 

From the above discussion, it can easily be seen that inside-out disk growth would always have the effect of {\it suppressing} disk flaring in the total stellar population, if not completely removing it. We produced the equivalents of Fig.~\ref{fig:2} for other disks from the two simulation suites considered here. The flaring was always strongly reduced when the total population was considered, suggesting that our results are generic.

\subsection{Radial age gradients at high distance from the disk midplane}
\label{sec:sim}

From the top panels of Fig.~\ref{fig:2} it can be expected that morphologically defined thick disks, as in observations of edge-on galaxies, will show a decrease in age with increasing disk radius. To quantify this, in the bottom row of Fig.~\ref{fig:2} we show the mean age variation with radius for stellar samples at different distance from the disk midplane, as indicated by the color bar on the right. Slices in $|z|$ have thickness $\Delta|z|=0.1h_d$. 

We also plot the mean age of samples close to (triangles) and high above (squares) the disk midplane, to be associated with thin and thick disks defined morphologically. The two vertical ranges considered, $|z|/h_d<0.1$ and $0.3<|z|/h_d<1.0$, for the Milky Way would correspond to $|z|<0.3$~kpc and $0.9<|z|<3.0$~kpc, using a disk scale-length $h_d=3.0$~kpc.

A decrease of age with radius is found for the above morphologically defined thin disk in both models. However, a significant negative age gradient is also seen in the morphological thick disks. Over the radial extent of 4 scale-lengths shown in Fig.~\ref{fig:2}, we find a drop in mean age of $\sim3$~Gyr and $\sim4.5$~Gyr for the thick disks of Model1 and Model2, respectively.

\section{Implications for the Milky Way}

We can already look for evidence of our prediction in Milky Way observations.

\subsection{Inversion in radial [$\alpha$/Fe] and [Fe/H] gradients away from the disk midplane}

As discussed above, our results suggest that even at distances high above the disk midplane, mean stellar age should decrease with galactic radius. As good age estimates for large stellar samples are currently unavailable, we can instead look at measurements of stellar [$\alpha$/Fe] ratios. Because stars of different masses release chemical elements to the interstellar medium on different timescales, abundance ratios, such as [$\alpha$/Fe], can be good proxies for age (e.g., \citealt{matteucci90, chiappini97}).

Negative [$\alpha$/Fe] gradients at large $|z|$ were reported by \cite{boeche13b} and \cite{boeche14} in dwarfs and giants, respectively, studying data from the RAVE survey \citep{steinmetz06}. \cite{anders14} and \cite{hayden14} also showed a transition from a weakly positive (or flat) to a negative radial gradient in APOGEE [$\alpha$/M] measurements, as the distance from the Milky Way disk midplane was increased. 
Similar inversion of [$\alpha$/Fe] gradients has been also found in recent chemo-dynamical models (\citealt{mcm14, rahimi14}; Miranda et al. 2015, submitted).

Flaring of mono-age disks can result also in the inversion of metallicity gradients with increasing distance from the disk midplane, as younger metal-rich stars in the outer disk can reach high vertical distances. Observationally, this inversion has been found in a number of spectroscopic Galactic surveys (e.g., SEGUE -- \citealt{cheng12a}, RAVE -- \citealt{boeche13b}, APOGEE -- \citealt{anders14}) and in simulations \citep{mcm14, rahimi14}.

All of the above results are in agreement with the decline in age with radius found in this work.

\subsection{Flaring of younger stellar populations}

Studying APOGEE data, \citep{nidever14} showed that, at $1<|z|<2$~kpc from the disk plane, the high-[$\alpha$/Fe] sequence dominates the inner disk ($5<r<7$~kpc), while in the outer disk ($9<r<11$~kpc) only the low-[$\alpha$/Fe] sequence is present (see their Fig.~10). 

Flaring of younger stellar populations in the Milky Way have been reported in several works (e.g., \citep{feast14, kalberla14, carraro15}.

All these results are in agreement with the idea of mono-age flaring populations suggested here, which allow chemically- or age-defined thin-disk stars to reach large vertical distances in the outer disk. 

\subsection{Extended morphologically defined thick disks}

We are suggesting here that inside-out disk formation inevitably gives rise to centrally concentrated older/[$\alpha$/Fe]-enhanced populations, but at the same time - to comparable scale-lengths for the morphologically defined thin and thick disks. It can be seen in the middle row of Fig.~\ref{fig:2} that this expectation is borne out for both our models. While inside 3-4$h_d$ mono-age populations become steeper with increasing age, the surface density profiles (of stars of all ages) close to (triangles) and high above (squares) the disk midplane run mostly parallel. Moreover, beyond 3-4$h_d$ the thick disk profiles become flatter than the thin ones, i.e., extend further out.

A morphologically defined thick disk in the Milky Way with a scale-length larger than that of the thin disk has been reported by \citealt{robin96} using a compilation of different data sets, \cite{ojha01} using 2MASS data, and \cite{juric08} using SDSS data. This is well in line with external observations and explained by the idea presented here, resolving the apparent disagreement with the centrally concentrated [$\alpha$/Fe]-enhanced populations in the Galaxy reported by \cite{bensby11} and \cite{bovy12a}.

\section{Discussion}

We showed that in galactic disks formed inside-out, mono-age populations are well fitted by single exponentials and always flare. In contrast, when the total stellar density is considered, a sum of two exponentials is required for a good fit, resulting in thin and thick disks, which do not flare. This can be explained by the increase in scale-length of younger populations flaring at progressively larger radii. This scenario resolves the apparent contradictions that chemically (or age) defined thick disks are centrally concentrated, but structurally thick populations in both observations of edge-on galaxies and in the Milky Way, extend as much as, or even beyond, the thinner component.

The result that flared mono-age populations give rise to non-flared thin and thick disks is directly linked to the disk inside-out formation. If galactic disks formed with a constant scale-length as a function of time, then the flares of all mono-age groups would simply add up, making significant flaring of the total stellar population unavoidable. This is not seen in observations of edge-on galaxies. Our results, thus, land support to this widely accepted idea of inside-out formation of galactic disks.

We found similar results in two simulation suites, represented here by Model1 and Model2, using radically different simulation techniques. This suggests that our results are generic and should be found in all cosmological simulations, where inside-out disk growth takes place. 

While flaring has been seen as a general characteristic of disk heating by mergers, this conclusion can be linked to the initial conditions of preassembled disks typically used in such studies (allowing for a systematic exploration of parameter space). However, the results of this work suggest that heating by mergers can produce thick disks which do not flare, by considering the more realistic scenario of inside-out growing disks in a cosmological environment. It also implies that flaring in the total stellar population in observations of external galaxies cannot be used as a measure of the amount of mergers suffered by the host disk. Instead, flaring in narrow age groups should be searched for.

The phenomenon described here results from the interplay among the intensity of inside-out formation and the masses of infalling satellites and their frequency as functions of redshift. Further work is necessary to find observables which can differentiate between these processes. Our results may then form the basis for future high-precision surveys (e.g., Gaia in the Milky Way) to constrain disk formation mechanisms by quantifying the amount of mergers and gas inflow as a function of cosmic time.

\acknowledgments
We thank the anonymous referee for a helpful report.
We also thank M. Aumer for making his simulations available to us and for useful comments, as well as L. Athanassoula, A. Helmi, J. Binney, and H.-W. Rix for helpful discussions.


\begin{thebibliography}{}

\bibitem[\protect\astroncite{{Anders} et~al.}{2014}]{anders14}
{Anders}, F., {Chiappini}, C., {Santiago}, B.~X., et al.: 2014,
\newblock {\em \aap} {\bf 564}, A115

\bibitem[Aumer \& White(2013)]{aumer13a} 
Aumer, M., \& White, S.~D.~M.\ 2013, \mnras, 428, 1055

\bibitem[\protect\astroncite{{Aumer} et~al.}{2013}]{aumer13b}
{Aumer}, M., {White}, S.~D.~M., {Naab}, T., and {Scannapieco}, C.: 2013,
\newblock {\em \mnras} {\bf 434}, 3142

\bibitem[Aumer et al.(2014)]{aumer14} 
Aumer, M., White, S.~D.~M., \& Naab, T.\ 2014, \mnras, 441, 3679

\bibitem[\protect\astroncite{{Bensby} et~al.}{2011}]{bensby11}
{Bensby}, T., {Alves-Brito}, A., {Oey}, M.~S., et al..: 2011,
\newblock {\em \apjl} {\bf 735}, L46

\bibitem[\protect\astroncite{{Brook} et~al.}{2012}]{brook12}
{Brook}, C.~B., {Stinson}, G.~S., {Gibson}, et al.: 2012,
\newblock {\em \mnras} {\bf 426}, 690

\bibitem[\protect\astroncite{{Boeche} et~al.}{2013}]{boeche13b}
{Boeche}, C., {Siebert}, A., {Piffl}, T., et al.: 2013,
\newblock {\em \aap} {\bf 559}, A59

\bibitem[Boeche et al.(2014)]{boeche14} 
Boeche, C., Siebert, A., Piffl, T., et al.\ 2014, \aap, {\bf 568}, AA71 

\bibitem[\protect\astroncite{{Bournaud} et~al.}{2009}]{bournaud09}
{Bournaud}, F., {Elmegreen}, B.~G., and {Martig}, M.: 2009,
\newblock {\em \apjl} {\bf 707}, L1

\bibitem[\protect\astroncite{{Bovy} et~al.}{2012}]{bovy12a}
{Bovy}, J., {Rix}, H.-W., {Liu}, C., et al.: 2012,
\newblock {\em \apj} {\bf 753}, 148

\bibitem[\protect\astroncite{{Burstein}}{1979}]{burstein79}
{Burstein}, D.: 1979,
\newblock {\em \apj} {\bf 234}, 829

\bibitem[Carraro et al.(2015)]{carraro15} 
Carraro, G., V{\'a}zquez, R.~A., Costa, E., Ahumada, J.~A., 
\& Giorgi, E.~E.\ 2015, \aj, {\bf 149}, 12 

\bibitem[\protect\astroncite{{Cheng} et~al.}{2012a}]{cheng12b}
{Cheng}, J.~Y., {Rockosi}, C.~M., {Morrison}, H.~L., et al.: 2012a,
\newblock {\em \apj} {\bf 752}, 51

\bibitem[\protect\astroncite{{Cheng} et~al.}{2012b}]{cheng12a}
{Cheng}, J.~Y., {Rockosi}, C.~M., {Morrison}, H.~L., et al.: 2012b,
\newblock {\em \apj} {\bf 746}, 149

\bibitem[Chiappini et al.(1997)]{chiappini97} 
Chiappini, C., Matteucci, F., \& Gratton, R.\ 1997, \apj, {\bf 477}, 765

\bibitem[\protect\astroncite{{Comer{\'o}n} et~al.}{2011}]{comeron11}
{Comer{\'o}n}, S., {Elmegreen}, B.~G., {Knapen}, J.~H., et al.: 2011,
\newblock {\em \apj} {\bf 741}, 28

\bibitem[\protect\astroncite{{Comer{\'o}n} et~al.}{2012}]{comeron12}
{Comer{\'o}n}, S., {Elmegreen}, B.~G., {Salo}, H., et al.: 2012,
\newblock {\em \apj} {\bf 759}, 98

\bibitem[Feast et al.(2014)]{feast14} 
Feast, M.~W., Menzies, J.~W., Matsunaga, N., \& Whitelock, P.~A.\ 2014, \nat, {\bf 509}, 342 

\bibitem[\protect\astroncite{{de Grijs}}{1998}]{degrijs98}
{de Grijs}, R.: 1998,
\newblock {\em \mnras} {\bf 299}, 595

\bibitem[\protect\astroncite{{Gilmore} and {Reid}}{1983}]{gilmore83}
{Gilmore}, G. and {Reid}, N.: 1983,
\newblock {\em \mnras} {\bf 202}, 1025

\bibitem[Guedes et al.(2011)]{guedes11} 
Guedes, J., Callegari, S., Madau, P., \& Mayer, L.\ 2011, \apj, {\bf 742}, 76 

\bibitem[House et al.(2011)]{hause11} 
House, E.~L., Brook, C.~B., Gibson, B.~K., et al.\ 2011, \mnras, {\bf 415}, 2652 

\bibitem[Hayden et al.(2014)]{hayden14} 
Hayden, M.~R., Holtzman, J.~A., Bovy, J., et al.\ 2014, \aj, {\bf 147}, 116 

\bibitem[Haywood et al.(2013)]{haywood13} 
Haywood, M., Di Matteo, P., Lehnert, M.~D., Katz, D., \& G{\'o}mez, A.\ 2013, \aap, {\bf 560}, 109

\bibitem[Juri{\'c} et al.(2008)]{juric08} 
Juri{\'c}, M., Ivezi{\'c}, {\v Z}., Brooks, A., et al.\ 2008, \apj, {\bf 673}, 864 

\bibitem[Kalberla et al.(2014)]{kalberla14} 
Kalberla, P.~M.~W., Kerp, J., Dedes, L., \& Haud, U.\ 2014, \apj, {\bf 794}, 90

\bibitem[Kazantzidis et al.(2008)]{kazantzidis08} 
Kazantzidis, S., Bullock, J.~S., Zentner, A.~R., Kravtsov, A.~V., 
\& Moustakas, L.~A.\ 2008, \apj, 688, 254

\bibitem[\protect\astroncite{{Martig} et~al.}{2012}]{martig12}
{Martig}, M., {Bournaud}, F., {Croton}, D.~J., et al.: 2012,
\newblock {\em \apj} {\bf 756}, 26

\bibitem[\protect\astroncite{{Martig} et~al.}{2009}]{martig09}
{Martig}, M., {Bournaud}, F., {Teyssier}, R., and {Dekel}, A.: 2009,
\newblock {\em \apj} {\bf 707}, 250

\bibitem[\protect\astroncite{{Martig} et~al.}{2014}]{martig14a}
{Martig}, M., {Minchev}, I., and {Flynn}, C.: 2014,
\newblock {\em \mnras} {\bf 442}, 2474

\bibitem[Martig et al.(2014)]{martig14b} 
Martig, M., Minchev, I., \& Flynn, C.\ 2014, \mnras, {\bf 443}, 2452 

\bibitem[Matteucci \& Brocato(1990)]{matteucci90} 
Matteucci, F., \& Brocato, E.\ 1990, \apj, 365, 539

\bibitem[\protect\astroncite{{Minchev} et~al.}{2012}]{minchev12b}
{Minchev}, I., {Famaey}, B., {Quillen}, A.~C., {Dehnen}, W., {Martig}, M., and {Siebert}, A.: 2012,
\newblock {\em \aap} {\bf 548}, 127

\bibitem[Minchev et al.(2013)]{mcm13} 
Minchev, I., Chiappini, C., \& Martig, M.\ 2013, \aap, {\bf 558}, AA9

\bibitem[Minchev et al.(2014a)]{minchev14} 
Minchev, I., Chiappini, C., Martig, M., et al.\ 2014a, \apjl, {\bf 781}, 20 

\bibitem[\protect\astroncite{{Minchev} et~al.}{2014b}]{mcm14}
{Minchev}, I., {Chiappini}, C., and {Martig}, M.: 2014b,
\newblock {\em \aap} {\bf 572}, 92

\bibitem[\protect\astroncite{{Nidever} et~al.}{2014}]{nidever14}
{Nidever}, D.~L., {Bovy}, J., {Bird}, J.~C., et al.: 2014,
\newblock {\em \apj} {\bf 796}, 38

\bibitem[Ojha(2001)]{ojha01} 
Ojha, D.~K.\ 2001, \mnras, {\bf 322}, 426 

\bibitem[Pohlen et al.(2007)]{pohlen07} 
Pohlen, M., Zaroubi, S., Peletier, R.~F., \& Dettmar, R.-J.\ 2007, \mnras, 378, 594

\bibitem[\protect\astroncite{{Rahimi} et~al.}{2014}]{rahimi14}
{Rahimi}, A., {Carrell}, K., and {Kawata}, D.: 2014,
\newblock {\em Research in Astronomy and Astrophysics} {\bf 14}, 1406

\bibitem[\protect\astroncite{{Robin} et~al.}{1996}]{robin96}
{Robin}, A.~C., {Haywood}, M., {Creze}, M., et al.: 1996,
\newblock {\em \aap} {\bf 305}, 125

\bibitem[Ro{\v s}kar et al.(2010)]{roskar10} 
Ro{\v s}kar, R., Debattista, V.~P., Brooks, A.~M., et al.\ 2010, \mnras, {\bf 408}, 783 

\bibitem[\protect\astroncite{{Steinmetz} et~al.}{2006}]{steinmetz06}
{Steinmetz, M., Zwitter, T., Siebert}, et~al. 2006,
\newblock{AJ}, {\bf 132}, 1645

\bibitem[van der Kruit \& Searle(1982)]{vanderkruit82} 
van der Kruit, P.~C., \& Searle, L.\ 1982, \aap, 110, 61 

\bibitem[Villalobos \& Helmi(2008)]{villalobos08} 
Villalobos, {\'A}., \& Helmi, A.\ 2008, \mnras, 391, 1806

\bibitem[Scannapieco et al.(2009)]{scannapieco09} Scannapieco, C., 
White, S.~D.~M., Springel, V., \& Tissera, P.~B.\ 2009, \mnras, {\bf 396}, 696

\bibitem[Springel et al.(2008)]{springel08} Springel, V., Wang, 
J., Vogelsberger, M., et al.\ 2008, \mnras, {\bf 391}, 1685

\bibitem[Stinson et al.(2013)]{stinson13} 
Stinson, G.~S., Bovy, J., Rix, H.-W., et al.\ 2013, \mnras, {\bf 436}, 625 

\bibitem[\protect\astroncite{{Tsikoudi}}{1979}]{tsikoudi79}
{Tsikoudi}, V.: 1979,
\newblock {\em \apj} {\bf 234}, 842

\bibitem[\protect\astroncite{{Yoachim} and {Dalcanton}}{2006}]{yoachim06}
{Yoachim}, P. and {Dalcanton}, J.~J.: 2006,
\newblock {\em \aj} {\bf 131}, 226

\end{thebibliography}
\end{document}